\documentclass[twocolumn]{aa}
\usepackage{natbib}
\usepackage{graphicx}
\usepackage{txfonts}

\defcitealias{landibommier94}{Paper~I}

\begin{document}

\title{Generalized $\sqrt{\epsilon}$-law}
\subtitle{The role of unphysical source terms in resonance line polarization
transfer and its importance as an additional test of NLTE radiative transfer
codes.}

\author{J. \v{S}t\v{e}p\'an\inst{1}\fnmsep\inst{2}\and V. Bommier\inst{2}}

\offprints{J. \v{S}t\v{e}p\'an}

\institute{Astronomical Institute, Academy of Sciences of the Czech
           Republic, 251\,65 Ond\v{r}ejov, Czech Republic\\
           \email{stepan@asu.cas.cz}
           \and
           LERMA, Observatoire de Paris -- Meudon, CNRS UMR 8112, 5,
           Place Jules Janssen, 92195 Meudon Cedex, France\\
           \email{[jiri.stepan;V.Bommier]@obspm.fr}
          }

\date{Received 4 October 2006 / Accepted 26 March 2007}

\abstract
      {A derivation of a generalized $\sqrt{\epsilon}$-law for
       nonthermal collisional rates of excitation by charged perturbers is
       presented.}
      {Aim of this paper is to find a more general
       analytical expression for a surface value of the source function
       which can be used as an addtional tool for verification of the
       non-LTE radiative transfer codes.}
      {Under the
       impact approximation hypothesis, static, one-dimensional, plane-parallel
       atmosphere, constant magnetic field of arbitrary strength and
       direction, two-level atom model with unpolarized lower level and
       stimulated emission neglected, we introduce the
       unphysical terms 
       into the equations of statistical equilibrium and
       solve the appropriate non-LTE integral equations.}
      {We derive a new analytical condition for the surface values of the 
       source
       function components expressed in the basis of irreducible
       spherical tensors.}
       {}

   
\maketitle
   

\section{Introduction}

In the series of papers of
\citet{landibommier91a,landibommier91b},
\citet{landibommier94} (from now on referenced as \citetalias{landibommier94}),
the general formalism of resonance line polarization scattering for
a two-level atom has been developed.
The \mbox{non-LTE} problem of the 2nd~kind for an
arbitrary magnetic field,
three-dimensional geometry of the medium and arbitrary irradiation by external
sources has been discussed. The effect of inelastic collisions with
charged perturbers has been considered for the particular case of a relative 
Maxwellian velocity distribution.

\citetalias{landibommier94}
analysed the analytical
properties of the solutions in the particular case of a one-dimensional, 
semi-infinite, static atmosphere with a constant magnetic field of arbitrary
strength and direction and assuming zero external irradiation of the 
atmosphere.
They derived a generalization of the well known $\sqrt{\epsilon}$-law
\citep[e.g.][]{avrett65,mihalas70,hubeny87}
for the case of polarized radiation and extended the previous results of
\citet{ivanov90}
who studied scattering in a non-magnetic
regime.

In most cases of practical interest the polarization degree
is rather small. The purpose of this paper is to find a new analytical solution
of the non-LTE problem in unphysical conditions in order to better verify
the accuracy of the polarized radiation transfer codes. This is done by introduction
of an unphysical source term in the polarization into the equations
of statistical equilibrium.
Such a generalization can be useful in testing the
accuracy of the radiative transfer codes whose purpose is to deal with the
non-thermal collisional processes (for instance in the impact polarization
studies of solar flares).

Following the approach of the papers quoted above, we adopt the formalism of
density matrix in the representation of irreducible
tensorial operators
\citep[e.g.][]{fano57}.
We consider the lower level with total angular momentum $j$ to be unpolarized.
This level is completely described by the overall population which is
set to 1 for normalization reasons.
The upper level with angular momentum $j'$ is described by the multipole
components of the statistical tensor $\rho^K_Q$.
Coherences between different levels $j$ and $j'$
are neglected but coherences between Zeeman sublevels of level $j'$ are
in general taken into account. The calculation is performed in the Wien
limit of line frequency whose assumption makes it possible to neglect
stimulated emission effect, and to preserve the linearity of the
\mbox{non-LTE} problem.


\section{Equations of statistical equilibrium}

\begin{figure}
 \centering
 \includegraphics[width=\columnwidth]{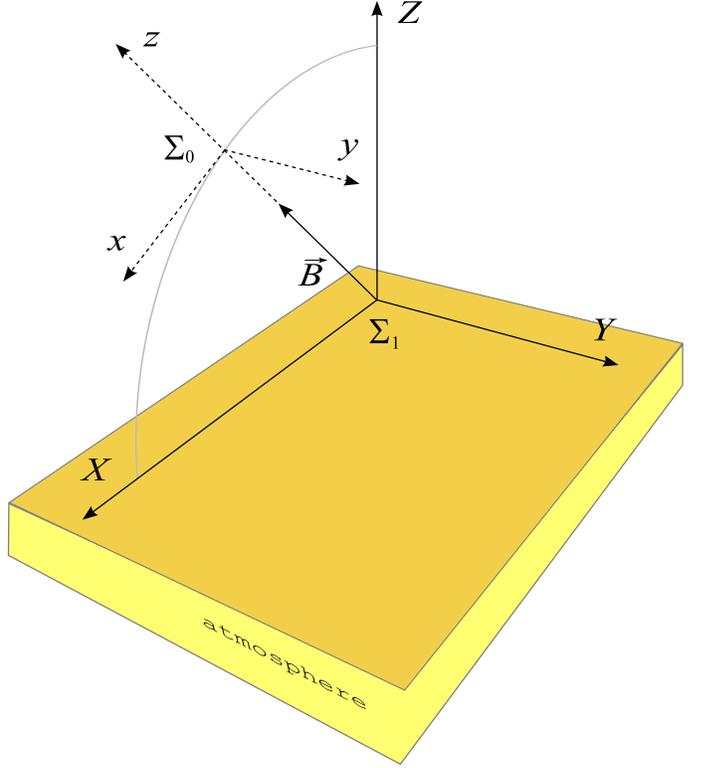} 
    \caption{The reference frame $\Sigma_1$ has its
        $Z$-axis oriented vertically with respect to the atmosphere, while
        the $z$-axis of the reference frame $\Sigma_0$ is parallel to the
        direction of magnetic field $\vec{B}$. The axes $X$ and $x$ lie
        in the same plane defined by $Z$-axis and $\vec{B}$; the axes
        $Y$ and $y$ are defined to complement the right-handed orthogonal
        coordinate systems.
            }
      \label{fig1}
\end{figure}

The suitable coordinate system $\Sigma_0$ for atomic state description
is the one with the $z$-axis directed along the magnetic
field (see Figure~\ref{fig1}).

Radiative rate contributions to the evolution of statistical
operator $\rho^K_Q$ are given by
\citep{landi85}
\begin{eqnarray}
  \left[ \frac{{\rm d}\rho^K_Q}{{\rm d}t} \right]_{\rm RAD}&=&
  -{\rm i}A_{j'j}\Gamma Q\rho^K_Q
  -A_{j'j}\rho^K_Q\nonumber\\
  &+&\frac {w^{(K)}_{j'j}(-1)^Q}{\sqrt{2j'+1}}B_{jj'}
  \overline{J}^K_{-Q}.
  \label{eq1}
\end{eqnarray}
In this equation $A_{j'j}$ ($B_{jj'}$) is the Einstein coefficient of
spontaneous emission (absorption) from level $j'$ ($j$) to level $j$ ($j'$).
$\Gamma=2\pi g_{j'}\nu_{\rm L}/A_{j'j}$ with $g_{j'}$ being the Land\'e
factor of the level $j'$ and $\nu_{\rm L}$ is the Larmor frequency.
The transition-dependent numerical factor $w^{(K)}_{j'j}$ 
has been defined by
\citet{landi84}
as have 
the irreducible components of the mean radiation tensor $\overline{J}^K_Q$.
Besides the radiative rates, collisional rates have to be considered in the
statistical equilibrium, because the source of radiation in a semi-infinite 
atmosphere is the collisional excitation followed by radiative de-excitation. 
Thus, the source term of the radiative transfer equation originates in the 
inelastic collision effect. As the purpose of the present paper is to consider 
unphysical source terms in the non-zero ranks $(K,Q)$ of the irreducible tensorial 
operator basis $T^K_Q$, we will introduce an unphysical $(K,Q)$-dependence to the 
inelastic collisional rates of the statistical equilibrium equation below. The 
purpose here is not to thus describe anisotropic collisions, which would require a 
proper formalism that is out of the scope of the present paper (see, for instance, 
\citet{landi04}
for a two-level atom, and \citet{derouich06}, 
for polarization transfer rates in a multi-level atom due to isotropic collisions). 
We introduce as usual the depolarizing rate due to isotropic elastic 
collisions. Thus, the contribution of collisional rates reads
\begin{equation}
  \left[ \frac{{\rm d}\rho^K_Q}{{\rm d}t} \right]_{\rm COLL}=
  (C_{jj'})^K_Q
  -(C^{\rm R}_{j'j})^K_Q\rho^K_Q
  -D^{(K)}\rho^K_Q.
  \label{eq2}
\end{equation}
The terms $(C_{jj'})^K_Q$ and $(C^{\rm R}_{j'j})^K_Q$ on the right-hand
side of equation (\ref{eq2}) are the multipole components of collisional
rates of excitation and relaxation respectively. 
$D^{(K)}$ is the
depolarization rate due to elastic collisions.\footnote{This process
cannot change a total population of the level. Therefore it is
always $D^{(0)}=0$.
We take formally into account only
the depolarization rate $D^{(K)}$ to use a formalism
coherent with the previous papers.
A general treatment of physically more relevant transfer of multipole components
of the upper level is out of scope of this paper.}

The radiative and collisional rates can be added under the impact
approximation hypothesis
\citep{bommier91}
${\rm d}\rho^K_Q/{\rm d}t=
[{\rm d}\rho^K_Q/{\rm d}t]_{\rm RAD}+
[{\rm d}\rho^K_Q/{\rm d}t]_{\rm COLL}$.
Using the equations (\ref{eq1}),
(\ref{eq2}), and the condition for static atmosphere,
${\rm d}\rho^K_Q/{\rm d}t=0$, we obtain the equations of
statistical equilibrium
\begin{eqnarray}
  [{\rm i}A_{j'j}\Gamma Q+A_{j'j}+(C^{\rm R}_{j'j})^K_Q+D^{(K)}]
  \rho^K_Q\nonumber\\
  =\frac {w^{(K)}_{j'j}(-1)^Q}{\sqrt{2j'+1}}B_{jj'}\overline{J}^K_{-Q}
  +(C_{jj'})^K_Q.
  \label{eq3}
\end{eqnarray}
By applying the relation between Einstein coefficients for spontaneous
emission and absorption,
\begin{equation}
  B_{jj'}=\frac{2j'+1}{2j+1}\frac{c^2}{2h\nu_0^3}A_{j'j},
\end{equation}
and dividing the formula (\ref{eq3}) by $A_{j'j}$,
we obtain the equation 
\begin{eqnarray}
  (1+\epsilon^K_Q+\delta^{(K)}+{\rm i}\Gamma Q)\rho^K_Q
  \nonumber\\
 =(-1)^Qw^{(K)}_{j'j}\overline{J}^K_{-Q}
 \frac{\sqrt{2j'+1}}{2j+1}\frac{c^2}{2h\nu_0^3}
  +\frac{(C_{jj'})^K_Q}{A_{j'j'}}.
  \label{eq4}
\end{eqnarray}
One can introduce the dimensionless parameter of the depolarization rate
\begin{equation}
  \delta^{(K)}=\frac{D^{(K)}}{A_{j'j}},
\end{equation}
and the irreducible tensor which plays the role of generalized photon
destruction probability
\begin{equation}
  \epsilon^K_Q=\frac{(C^{\rm R}_{j'j})^K_Q}{A_{j'j'}}.
  \label{eq20}
\end{equation}
If the relation $(C^{\rm R}_{j'j})^K_Q\neq 0$ is satisfied
we may define the quantity
\begin{equation}
  B^{(KQ)}=\frac{2h\nu_0^3}{c^2}\frac{2j+1}{\sqrt{2j'+1}}
  \frac{(C_{jj'})^K_Q}{(C^{\rm R}_{j'j})^K_Q}.
\end{equation}
It is easy to show (see below) that in the particular case of a Maxwellian
velocity distribution
of colliders the relation $B^{(00)}=B_{\rm P}$ is satisfied, where
$B_{\rm P}$ is the Planck function in the Wien limit at given temperature.
Using the definition of irreducible components of the two-level source
function
\citepalias[cf.][]{landibommier94}
\begin{equation}
  S^K_Q=\frac{2h\nu_0^3}{c^2}\frac{2j+1}{\sqrt{2j'+1}}\rho^K_Q,
\end{equation}
we obtain the statistical equilibrium equations in the compact form
\begin{eqnarray}
  (1+\epsilon^K_Q+\delta^{(K)}+{\rm i}\Gamma Q)S^K_Q\nonumber\\
  =w^{(K)}_{j'j}(-1)^Q\overline{J}^K_{-Q}
 +\epsilon^K_Q B^{(KQ)}.
\end{eqnarray}


\section{Solution of the Wiener-Hopf equations}
  
From now on we reduce our analysis to the case of semi-infinite,
plane-parallel geometry with constant magnetic field along
the atmosphere. The velocity distribution and volume density of
colliders is also constant along the atmosphere but it is
in general non-thermal. The only position coordinate is the
common line optical depth $\tau$.

Following the procedure of \citetalias{landibommier94}
a formal solution of radiative transfer equation is substituted
into the definition of tensor $\overline{J}^K_Q$; after that we
obtain a set of integral Wiener--Hopf equations of the
2nd~kind,
\begin{eqnarray}
  (1+\epsilon^K_Q+\delta^{(K)}+{\rm i}\Gamma Q)S^K_Q(\tau)\nonumber\\
  =\sum_{K'Q'}\int_0^\infty \widetilde{K}_{KQ,K'Q'}(\tau,\tau')
  S^{K'}_{Q'}(\tau'){\rm d}\tau'+\epsilon^K_Q B^{(KQ)},
  \label{eq21}
\end{eqnarray}
which describe coupling of the tensors $\rho^K_Q(\tau)$ at different optical
depths via radiation.
Several important properties of kernels $\widetilde{K}_{KQ,K'Q'}(\tau,\tau')$
have been discussed by \citet{landibommier90}
and in \citetalias{landibommier94}
Using their indexing notation one can
rewrite the equation (\ref{eq21}) in the shorthanded form
\begin{equation}
  a_i S_i(\tau)=\sum_{j}\int_0^\infty K_{ij}(|\tau-\tau'|)S_j(\tau')
  {\rm d}\tau'+b_i,
  \label{eq8}
\end{equation}
with
\begin{eqnarray}
a_i&=&1+\epsilon^K_Q+\delta^{(K)}+{\rm i}\Gamma Q,\\
b_i&=&\epsilon^K_Q B^{(KQ)}.
\end{eqnarray}
The index $i$ in these expressions runs between the limits 1 and
$N$, where $N$ is the number of ${}^K_Q$-multipoles.
In the following we briefly repeat the derivation
performed by 
\citet{frisch75}
emphasizing the differences due to presence of $b_i$ terms.

Calculation of the derivative of (\ref{eq8}) with respect to
$\tau$, splitting the integral on the right-hand side into two parts,
multiplication of the equation
by $S_i(\tau)$, summation over index $i$, and finally integration
with respect to $\tau$ leads to the set of equations
\begin{eqnarray}
  \sum_i a_i\int_0^{\infty}S_i(\tau)\frac{{\rm d}S_i(\tau)}{{\rm d}\tau}
  {\rm d}\tau\nonumber\\
  =\sum_{i,j}S_j(0)\int_0^{\infty}K_{ij}(\tau)S_i(\tau){\rm d}\tau
  \nonumber\\
  +\sum_{i,j}\int_0^{\infty}{\rm d}\tau S_i(\tau)
  \int_0^{\infty}{\rm d}\tau'K_{ij}(|\tau'-\tau|)
  \frac{{\rm d}S_j(\tau')}{{\rm d}\tau'}.
  \label{eq9}
\end{eqnarray}
The left-hand side of (\ref{eq9}) is easily evaluated as
\begin{equation}
  \frac 12 \sum_i a_i \left[S_i(\infty)^2-S_i(0)^2\right]
\end{equation}
The first term on the right-hand side of (\ref{eq9}) is evaluated using 
the kernels symmetry $K_{ij}(t)=K_{ji}(t)$ 
and the equation (\ref{eq8}), so that we obtain
\begin{equation}
  \sum_i S_i(0)\left[a_iS_i(0)-b_i\right],
\end{equation}
while the second term equals
\begin{equation}
  \frac 12\sum_i a_i\left[S_i(\infty)^2-S_i(0)^2\right]-
  \sum_i b_i\left[S_i(\infty)-S_i(0)\right].
\end{equation}
We put these results into (\ref{eq9}) to get
\begin{equation}
  \sum_i a_i S_i(0)^2=\sum_i b_i S_i(\infty).
  \label{eq40}
\end{equation}

Calculation of the limit $\tau\to\infty$ of both sides of the equation
(\ref{eq8}) leads to the set of linear algebraic equations
for the components of source function tensor in the infinite depth:
\begin{equation}
  \sum_j \left[a_j\delta_{ij}-
  \int_{-\infty}^{\infty}K_{ij}(t){\rm d}t\right]S_j(\infty)=b_i.
  \label{eq10}
\end{equation}
We can solve these equations and write
\begin{equation}
  \vec{S}(\infty)=\vec{L}^{-1}\vec{b},
  \label{eq41}
\end{equation}
where $\vec{S}$ is the formal vector of $S_i$ components, $\vec{b}$ is the
formal vector of $b_i$ components, and the elements of matrix $\vec{L}$
are defined by relation
\begin{equation}
  \{\vec{L}\}_{ij}=a_j\delta_{ij}-
  \int_{-\infty}^{\infty}K_{ij}(t){\rm d}t.
  \label{eq42}
\end{equation}
Establishing a new matrix
$\vec\ell=\vec L^{-1}$ and substituting (\ref{eq41}) into (\ref{eq40})
leads to the generalized form of the $\sqrt{\epsilon}$-law
\begin{equation}
  \sum_i a_i S_i(0)^2=\sum_{i,j} b_i b_j\ell_{ij}.
  \label{eq5}
\end{equation}


\section{Particular solutions}

Setting the special conditions for magnetic field and collisional
rates, one recovers the less general but more common and explicit
formulations of the $\sqrt{\epsilon}$-law than the one given by
(\ref{eq5}). In the following sections we will verify this
result in the limiting conditions assumed in recent papers and 
we will analyse the simple examples of non-thermal collisional excitation.


\subsection{Maxwellian velocity distribution of colliders}

In the case of Maxwellian velocity distribution of colliders,
relaxation rates of all multipole components $\rho^K_Q$ are the same:
\begin{equation}
  (C^{\rm R}_{j'j})^K_Q=
  C^{\rm R}_{j'j},
  \label{eq55}
\end{equation}
where $C^{\rm R}_{j'j}$ is the usual relaxation rate
for collisional deexcitation from $j'$ to $j$. For excitation rates
one has
\begin{equation}
  (C_{jj'})^K_Q=\frac{C_{jj'}}{\sqrt{2j'+1}}\delta_{K0}\delta_{Q0},
  \label{eq56}
\end{equation}
where the factor $(2j'+1)^{-1/2}$ has been introduced to make a connection
with the usual collisional rate $C_{jj'}$ of standard unpolarized theory.
In this isotropic case, there is no collisional excitation of higher ranks of
density matrix. From the assumption of thermodynamic equilibrium one has
\begin{equation}
  \frac{C_{jj'}}{C^{\rm R}_{j'j}}=\frac{2j'+1}{2j+1}
  {\rm e}^{-h\nu_0/k_{\rm B}T},
\end{equation}
where $k_{\rm B}$ stands for the Boltzmann constant and $T$ for
a temperature of the atmosphere.
From (\ref{eq55}) and (\ref{eq20}) it is evident that
$\epsilon^K_Q=\epsilon$ for all possible $K$ and $Q$, where $\epsilon$ is
the common photon destruction probability. Further
\begin{equation}
  B^{(KQ)}=B_{\rm P}\delta_{K0}\delta_{Q0}.
\end{equation}
Substituting the rates (\ref{eq55}) and (\ref{eq56}) into
the formula (\ref{eq42}) and employing the general identity
$\int_{-\infty}^{\infty}K_{i1}(t){\rm d}t=\delta_{i1}$
\citepalias[see][]{landibommier94}
together with
$b_i=\delta_{i1}$,
we recover form (\ref{eq5}) the formula (16) of the previously cited paper:
\begin{equation}
  \sum_{KQ}(1+\epsilon+\delta^{(K)}+{\rm i}\Gamma Q)[S^K_Q(0)]^2=
  \epsilon B^2_{\rm P}.
  \label{eq32}
\end{equation}

Assuming that there is zero magnetic field, i.e. $\Gamma=0$, the source
function tensor reduces due to symmetry reasons to the two non-vanishing
components $S^0_0$ and $S^2_0$ in the reference frame $\Sigma_1$.
This reference frame is suitable for descriptions of the atomic system under
these conditions, so that we may identify $\Sigma_0\equiv\Sigma_1$, with
$X$ and $Y$ axes oriented arbitrary in the plane parallel to
atmospheric surface.
Further, assuming that there is no depolarization of the upper
level ($\delta^{(K)}=0$), we realize from (\ref{eq32}):
\begin{equation}
  \sqrt{\left[S^0_0(0)\right]^2+\left[S^2_0(0)\right]^2}=
  \sqrt{\frac{\epsilon}{1+\epsilon}}B_{\rm P}=\sqrt{\epsilon'}B_{\rm P},
  \label{eq6}
\end{equation}
which is the same result derived in different notation
by \citet{ivanov90}.
For simplicity
the common alternative to the photon destruction probability
has been introduced:
$\epsilon'=\epsilon/(1+\epsilon)$.

If depolarization of the upper level is high enough to destroy
atomic level polarization ($\delta^{(K)}\to\infty$ for $K>0$), or
the upper level is unpolarizable,
the common $\sqrt{\epsilon}$-law for scalar radiation is recovered,
\begin{equation}
  S^0_0(0)=\sqrt{\frac{\epsilon}{1+\epsilon}}B_{\rm P}=
  \sqrt{\epsilon'}B_{\rm P}.
\end{equation}


\subsection{Anisotropic alignment (de)excitation}

The relation
$\epsilon^K_Q=\epsilon$ is not in general satisfied for all the
multipoles because the relaxation of the $\rho^K_Q$ state
depends on the velocity distribution of colliders.
In the following text we will neglect the effects of magnetic field.

Let us assume an example of a relative velocity distribution of particles
that is axially symmetric with the axis of symmetry parallel to the
vertical of the atmosphere (so that it is as in the former
case $\Sigma_0\equiv\Sigma_1$)
and that the collisional interaction can be fully described by only the
first two even multipole components of this distribution.
Thanks to these assumptions the only non-vanishing excitation collisional
rates are $(C_{jj'})^0_0$ and $(C_{jj'})^2_0$,
the relaxation rates $(C^{\rm R}_{j'j})^0_0$ and
$(C^{\rm R}_{j'j})^2_0$ and for the same reasons the only non-zero source
function components are $S^0_0$ and $S^2_0$.

An explicit evaluation of the integrals of kernels
$\int_{-\infty}^{\infty}\widetilde{K}_{KQ,K'Q'}(\tau,\tau'){\rm d}\tau'$
under these conditions shows that the only non-zero ones 
are given by (A5) and (A12) of
\citet{landibommier91b}.
In our notation they read
\begin{eqnarray}
  \int_{-\infty}^{\infty}\widetilde{K}_{00,00}(\tau,\tau'){\rm d}\tau'&=&1,\\
  \int_{-\infty}^{\infty}\widetilde{K}_{20,20}(\tau,\tau'){\rm d}\tau'&=&
  \frac{7}{10}W_2,
\end{eqnarray}
with $W_2=(w^{(2)}_{j'j})^2$.
Substituting these results into (\ref{eq5}) we see that
\begin{eqnarray}
  (1+\epsilon^0_0)(S^0_0)^2+(1+\epsilon^2_0)(S^2_0)^2\nonumber\\
  =\epsilon^0_0 (B^{(00)})^2+
  \frac{(\epsilon^2_0 B^{(20)})^2}{1+\epsilon^2_0-\frac{7}{10}W_2}.
  \label{eq30}
\end{eqnarray}
  
To check the validity of polarized radiative transfer codes, it is
advantageous if one can verify that the transfer of higher ranks
of the radiation tensor is accurate enough. In the realistic scattering
polarization models the polarization degree does not exceed a few percent
so that $|S^0_0(0)|\gg |S^K_Q(0)|$. By setting arbitrary
(even unphysical) collisional rates it is possible to verify transfer
codes in conditions with $|S^0_0|\ll |S^K_Q|$.

To privilege transfer in higher ranks of the radiation tensor
one can artificially suppress the excitation rate $(C_{jj'})^0_0$.
In the extremal case one can set $(C_{jj'})^0_0\to 0$.
The easiest way to do this is the formal interchange of the role of
excitation rates of population and alignment, i.e.
$(C_{jj'})^0_0\leftrightarrow (C_{jj'})^2_0$ of the original Maxwellian
velocity distribution:
\begin{equation}
  (C_{jj'})^0_0=0,\quad (C_{jj'})^2_0=\frac{C_{jj'}}{\sqrt{2j'+1}}
  \label{eq31}
\end{equation}
(no collisional excitation to upper level population) and the relaxation
rates set to the Maxwellian ones
\begin{equation}
  (C^{\rm R}_{j'j})^0_0=(C^{\rm R}_{j'j})^2_0=C^{\rm R}_{j'j}.
\end{equation}
In this case we have
\begin{equation}
  B^{(00)}=0,\quad B^{(20)}=B_{\rm P},
\end{equation}
and again
\begin{equation}
  \epsilon^0_0=\epsilon^2_0=\epsilon.
\end{equation}
Substituting this into (\ref{eq30}) we find out the
$\sqrt{\epsilon}$-law in the form
\begin{equation}
  \sqrt{[S^0_0(0)]^2+[S^2_0(0)]^2}=
  \frac{\epsilon'}{\sqrt{1-\frac {7}{10}W_2(1-\epsilon')}}B_{\rm P}.
  \label{eq7}
\end{equation}
The particular collisional
rates (\ref{eq31}) are in fact arbitrary and have been chosen
to obtain a formula similar to the one of the Maxwellian distribution case.

This relation is useful to test polarized radiative transfer codes, because in 
this unphysical case $S^2_0(0)$ is the largest term, unlike the physical 
case where the largest term is $S^0_0(0)$ and $S^2_0(0)$ is only a few percent of 
it. By applying Eq.~(\ref{eq7}) the test is much more sensitive to the polarization, and 
the polarization is better tested. We have thus successfully tested a multilevel 
non-LTE radiative transfer code that we are developing, but this code and its 
results are the subjects of a forthcoming paper.


\section{Conclusions}

We have derived a more general formulation of the
so-called $\sqrt{\epsilon}$-law of radiation transfer.
This analytical condition couples the
value of source function tensor of a two-level atom with other physical
properties of the atmosphere.
The simplest result obtained in conditions of a non-magnetic, isothermal,
plane-parallel, semi-infinite atmosphere with thermal velocity distribution
of particles and unpolarized atomic levels
\citep[e.g.][]{mihalas70}
has been generalized by
\citet{ivanov90}
to account for scattering of
polarized radiation and polarized upper atomic level. Further generalizations
done in
\citetalias{landibommier94},
which account for a magnetic field of arbitrary
strength and direction, has been extended in the present paper to account
for non-thermal collisional interactions. It was done by introducing
the tensor of the photon destruction probability
$\epsilon^K_Q$ and by defining the function $B^{(KQ)}$.

The resulting formula (\ref{eq5}) reduces to the cases mentioned
above if the physical conditions become more symmetric. On the other hand,
situations with a high degree of perturbers velocity distribution anisotropy
and especially ones with unphysical collisional rates
result in a wide range of models which can be calculated both numerically
and analytically. Thus they offer new possibilities for
verification of the non-LTE radiation transfer codes.

\bibliographystyle{aa}
\bibliography{bibs}

\clearpage

\end{document}